\documentclass[9pt,twocolumn,twoside]{osajnl}
\journal{josab} % Choose journal (ao, aop, josaa, josab, ol, optica, pr)

\setboolean{shortarticle}{true}
\title{Analytical formulation of a high-power Yb-doped radiation balanced fiber laser}

\author[1,2,*]{Mostafa Peysokhan}
\author[1,2]{Brian Topper}
\author[1,2]{Esmaeil Mobini}
\author[1,2]{Arash Mafi}

\affil[1]{University of New Mexico, Physics \& Astronomy and Interdisciplinary Science, 210 Yale Blv NE, Albuquerque, NM 87131, USA}
\affil[2]{Center for High Technology Materials, University of New Mexico, 1313 Goddard St SE, Albuquerque, NM 87106, USA }

\affil[*]{Corresponding author: mpeysokhan@unm.edu}

\begin{abstract}

We present an analytical solution for a radiation balanced fiber laser, in which no net heat is generated during lasing operation due to cooling by anti-Stokes fluorescence. The results are in excellent agreement with the numerical solutions. Using realistic values for fiber laser parameters, the analytical solutions presented are proposed as an economically desirable alternative to time-consuming, direct numerical calculations. 
\end{abstract}

\setboolean{displaycopyright}{true}

\begin{document}

\maketitle

\section{Introduction}
The first demonstration of glass fiber lasers by Snitzer~\cite{Koester:64} marked a milestone achievement in the pursuit of reliable and compact laser systems. Fiber lasers and amplifiers are the first choices for industries that demand a stable, compact, low maintenance, and inexpensive high power source of laser light. Nowadays, high-power Yb-doped double-cladding fiber lasers (YDCFL) are one of the primary sources of high-power radiation for industrial and directed energy applications~\cite{Richardson:10, Zervas}. Fiber lasers have succeeded in keeping up with the demand from both industrial and defense sectors seeking higher power outputs. As a result of this effort, fiber lasers can now deliver powers on the scale of a few kilowatts (kW)~\cite{Jeong:09, Yan, Xiao:12, jauregui2013high, Beier:17}. 

On account of the excellent performance displayed by Yb-doped high-power fiber lasers and amplifiers, this research area has received significant attention in recent decades. The ultimate goal is to achieve higher laser output, accompanied by excellent stability and efficiency. Temperature rise in the core of the fiber has been a barrier to acceptable stability and efficiency in high power operation~\cite{Richardson}. Liquid-forced cooling~\cite{karimi2018theoretical,beier2017single,Beier:18,Jauregui:12} has been employed to solve the perpetual excess heat issue. However, the demand for more efficient heat mitigation persists in the efforts to develop a new generation of devices.

One possible solution to the thermal issues is called the Radiation Balanced Laser (RBL) technique or Radiation Balancing~\cite{PhysRevLett.127.013903,knall2021radiation}. In this approach, the fiber core's temperature can be decreased directly by mitigating the heat density via anti-Stokes fluorescence cooling. S.~Bowman initially introduced the concept of an RBL in 1999~\cite{bowman1999lasers,Mobini:18}.The concept leverages the phenomena of solid-state laser cooling to yield a vibration-free method of heat mitigation. Laser cooling may occur if an ultra high purity rare-earth-doped solid is pumped at wavelength $\lambda_p$ such that $\lambda_p$ > $\lambda_{f}$  where $\lambda_{f}$ is the mean fluorescence wavelength of the active ions. In doing so, the high population of emitted photons with a shorter wavelength than the pump effectively extracts heat from the solid. For a gain medium to be receptive to laser cooling, not only must the material be ultra pure since impurities and defects will act as sources of parasitic heating~\cite{knall2021design,knall2018model}, but consideration must also be given to the chosen active ion relative to the phonon edge of the host material. Such requirements impose rigorous restrictions on the selection of gain media to be used for radiation balancing amplifiers and lasers.

After the observation of solid-state laser cooling in Yb-doped silica glass recently~\cite{mobini2020laser, knall2020experimental,peysokhan2020laser, Mansoor,topper2021laser,knall2021radiation,PhysRevLett.127.013903} the idea of radiation balancing in fiber lasers and amplifiers seems more realistic. Bowman's expression of the RBL formalism in a bulk material results in a concluding equation that does not consider any relationship between pump and signal powers. In fiber lasers and amplifiers, pump and signal powers are not independent. With the recent advances in optical cooling of solids, the absence of an analytical solution for heat generation in fiber lasers that can lead to an RBL situation is perceived more than in the past. This paper is an extension of our previous work on finding an analytical solution for fiber lasers. Here, we show an analytical solution for heat generation in the fiber. This approach can serve as a guide for modifying the laser parameters and their influence on contributing to the RBL situation, or detracting from it.
\section{Overview of Anti-Stokes Fluorescence}
As mentioned before, laser cooling is the basis of the radiation balancing technique. To thoroughly comprehend the idea of radiation balancing, we need to understand optical refrigeration first. We will start our discussion with anti-Stokes fluorescence cooling. We begin by considering all the physical processes involved in the heat generation and extraction mechanisms in order to express the net power density, $Q$.
%%%%%%%%%%%%%%%%%%%%%%%%%%%%%%%%%%%%%%%%%%%%%%%%%%%%%%%%%%%%%%%%%%%%%%%%%%%%%%%%%%%%%%%%%%%%%%
\begin{figure}[htbp]
\centering 
    \includegraphics[width=3 in]{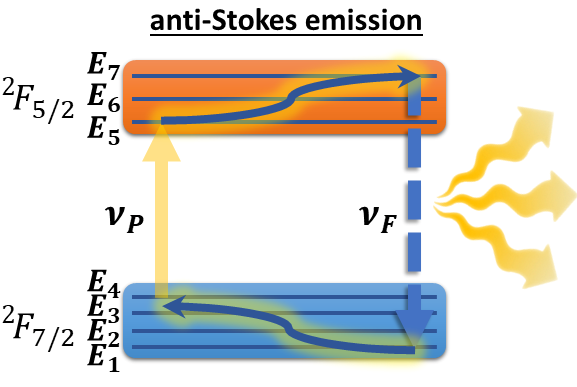}
\caption{Schematic diagram of anti-Stokes fluorescence cooling in a Yb-doped material with Yb$^{3+}$ energy levels.}
\label{Fig:RefInfTem}
\end{figure}
%%%%%%%%%%%%%%%%%%%%%%%%%%%%%%%%%%%%%%%%%%%%%%%%%%%%%%%%%%%%%%%%%%%%%%%%%%%%%%%%%%%%%%%%%%

While pumping the gain medium  at wavelength $\lambda_p$ with intensity $I_p$, some of the active ions are excited to an upper level, $N_{2}$. Spontaneous decay to the excited state down to the ground state occurs at the rate $N_{2}/\tau_{f}$. A portion of the input intensity also gets absorbed by impurities, accounted for by $\alpha_{b}$. Assuming that the sample is not too long, the ion density in the excited state ($N_{2}$) can be described by
%%%%%%%%%%%%%%%%%%%%%%%%%%%%%%%%%%%%%%%%%%%%%%%%%%%%%%%%%%%%%%%%%%%%%
\begin{align}
\label{Eq:Upperstatepop-1}
&\frac{dN_{2}}{dt}=\frac{I_{p}}{h\nu_{p}}\big(N_{0}\sigma_{a}(\lambda_{p})-N_{2}(\sigma_{a}(\lambda_{p})+\sigma_{e}(\lambda_{p}))\nonumber \\
&+\frac{I_{s}}{h\nu_{s}}\big(N_{0}\sigma_{e}(\lambda_{s})-N_{2}(\sigma_{e}(\lambda_{s})+\sigma_{a}(\lambda_{s})\big)-\frac{N_{2}}{\tau_{f}},
\end{align}
%%%%%%%%%%%%%%%%%%%%%%%%%%%%%%%%%%%%%%%%%%%%%%%%%%%%%%%%%%%%%%%% 
where the equation under steady state condition ($dN_2/dt=0$) leads to 
%%%%%%%%%%%%%%%%%%%%%%%%%%%%%%%%%%%%%%%%%%%%%%%%%%%%%%%%%%%%%%%%%%%%%
\begin{align}
\label{Eq:Upperstatepop-2}
&N_{2}=\frac{\beta_{p}i_{p}+\beta_{s}i_{s}}{1+i_{p}+i_{s}}N_{0},\quad\quad i_{i}=\frac{I_{i}}{I_{sat}(\lambda_{i})},\nonumber\\
&I_{sat}(\lambda_{i})=\frac{hc}{\lambda_{i}(\sigma_{a}(\lambda_{i})+\sigma_{e}(\lambda_{i}))\tau_{f}},\quad i\in\{p,s\}
\end{align}
%%%%%%%%%%%%%%%%%%%%%%%%%%%%%%%%%%%%%%%%%%%%%%%%%%%%%%%%%%%%%%%% 
where 
\begin{align}
\beta_i=\frac{\sigma_{a}(\lambda_i)}{\sigma_{a}(\lambda_i)+\sigma_{e}(\lambda_i)},\quad i\in\{p,s\} .
\end{align}
Here, $i_{p}$ and $i_{s}$ are the normalized pump and signal intensities, respectively, and $I_{s}$ is the saturation intensity. Since cooling samples are usually short, we may neglect the signal intensity ($i_{s}\approx 0$). In turn, the net power cooling density ($Q$) may be expressed as
%%%%%%%%%%%%%%%%%%%%%%%%%%%%%%%%%%%%%%%%%%%%%%%%%%%%%%%%%%%%%%%%%%%%%%%%%%%%%%%%%%%
\begin{align}
\label{Eq:Netcoolingdens-1}
Q(\lambda_{p})&=[N_{0}-\frac{N_{2}}{\beta_{p}}]\sigma_{a}(\lambda_{p}) I_{p}-\frac{ h c}{\lambda_{f}\tau_{r}} N_{2}+\alpha_{b} I_{p},
\end{align}
%%%%%%%%%%%%%%%%%%%%%%%%%%%%%%%%%%%%%%%%%%%%%%%%%%%%%%%%%%%%%%%%%%%%%%%%%%%%%%%%%%%%%%%%%%%%%%%%%%
where $\lambda_{f}$ is the mean fluorescence wavelength, $\alpha_{b}$ is the background absorption, and $\tau_{r}$ is the radiative lifetime.

In Eq.~(\ref{Eq:Netcoolingdens-1}), the first term represents the resonant absorption and is responsible for the heat generation. The second term originates from the anti-Stokes fluorescence and accounts for the heat extraction. Background absorption gives rise to the third term, contributing also to the heat generation in the material~\cite{mungan2003thermodynamics,MansoorSheik-bahae}.
The anti-Stokes fluorescence (ASF) emission responsible for the heat extraction in laser cooling can be understood by recalling that each excited atom in the upper level can decay from different sub-energy levels. If the average energy of each emitted photon is $\bar{E}_{ASF}=hc/\lambda_{f}$, then the net power density that escapes from the material over the radiative lifetime ($\tau_{r}$) would be $Q_{ASF}= (N_{2}hc/ \tau_{r}\lambda_{f})$.

The mean fluorescence wavelength ($\lambda_{f}$) is the average wavelength associated with an emitted photon. If we consider that $\phi(\nu)$ is the photon flux density, then the average energy of a photon that is emitted via ASF ($\bar{E}_{ASF}$) takes the following form
%%%%%%%%%%%%%%%%%%%%%%%%%%%%%%%%%%%%%%%%%%%%%%%%%%%%%%%%%%%%%%%%
\begin{align}
\label{Eq:MeaneFlourWavelength-flux}
\bar{E}_{ASF}=h\nu_{f}=h\frac{\int\phi(\nu)\nu d\nu}{\int\phi(\nu)d\nu}.
\end{align}
%%%%%%%%%%%%%%%%%%%%%%%%%%%%%%%%%%%%%%%%%%%%%%%%%%%%%%%%%%%%%%%

Considering the fact that $d\nu=-cd\lambda /\lambda^{2}$ and $(hc/\lambda) S(\lambda) = h \nu \phi(\nu)$, where $S(\lambda)$ is the spectral density, the mean fluorescence wavelength can be obtained from~\cite{bowman1999lasers,epstein2010optical}
%%%%%%%%%%%%%%%%%%%%%%%%%%%%%%%%%%%%%%%%%%%%%%%%%%%%%%%%%%%%%%%%
\begin{align}
\label{Eq:MeaneFlourWavelength}
\lambda_{f}=\frac{\int S(\lambda)\lambda d\lambda}{\int S(\lambda) d\lambda}.
\end{align}
%%%%%%%%%%%%%%%%%%%%%%%%%%%%%%%%%%%%%%%%%%%%%%%%%%%%%%%%%%%%%%%

Simplifying Eq. (\ref{Eq:Netcoolingdens-1}), we obtain the following equation for the net cooling power density,
%%%%%%%%%%%%%%%%%%%%%%%%%%%%%%%%%%%%%%%%%%%%%%%%%%%%%%%%%%%%%%%%%%%%%%%%%%%%%%%%%%%%%%%%%%%%%%%%%%%%  
\begin{align}
\label{Eq:Netcoolingdens-2}
Q(\lambda_{p})=(1-\eta_{ext}\frac{\lambda_{p}}{\lambda_{f}})\alpha_{r}(\lambda_{p})I_{p}+\alpha_{b} I_{p},
\end{align}
%%%%%%%%%%%%%%%%%%%%%%%%%%%%%%%%%%%%%%%%%%%%%%
where
%%%%%%%%%%%%%%%%%%%%%%%%%%%%%%%%%%%%%%%%%%%%%%
\begin{align}
\label{Eq:Extquantumeff-def}
&\eta_{ext}=\frac{\tau_f}{\tau_{r}},\quad\quad \tau_{f}=\eta_{e}\tau_{F},\nonumber\\
&\alpha_{r}(\lambda_{p})=[N_{0}-\frac{N_{2}}{\beta_{p}}]\sigma_{a}(\lambda_{p})=\frac{\alpha_{0}}{1+i_{p}},\quad\quad \alpha_{0}=N_{0}~\sigma_{a}(\lambda),
\end{align}
%%%%%%%%%%%%%%%%%%%%%%%%%%%%%%%%%%%%%%%%%%%%%%%%
and $\eta_{e}$ is the escape efficiency and $\tau_{f}$ is the modified fluorescence lifetime due to reabsorption. $\eta_{ext}$ is the external quantum efficiency that describes the amount of ASF that can leave the material. The population of photons not dedicated to the heat extraction via ASF ($(1-\eta_{ext})N_{2}$) go through either a non-radiative decay or a reabsorption process. The former is described by the escape efficiency ($\eta_{e}$) while the latter shows up in the internal quantum efficiency ($\eta_{q}$). One way to visualize the internal quantum efficiency is constructing two parallel channels through which the excited atom may return to a ground level, a radiative ($\tau_{r}$) channel and a non-radiative ($\tau_{nr}$) channel. The radiative part is that which contributes to heat extraction in Eq.(~\ref{Eq:Netcoolingdens-1}). 

Equation ~(\ref{Eq:Netcoolingdens-2}) may be written as a function of the pump intensity, $I_p$,  
%%%%%%%%%%%%%%%%%%%%%%%%%%%%%%%%%%%%%%%%%%%%%%%%%%%%%%%%%%%%%%%%%%%%%%%%%%%%%%%%%%%%%%%%%%%%%%%%%%%%  
\begin{align}
\label{Eq:Netcoolingdens-3}
Q(I_{p})=(1-\eta_{ext}\eta_{abs}\frac{\lambda_{p}}{\lambda_{f}})\alpha_{r}(I_{p})I_{p},\quad\quad 
\eta_{abs}=\frac{\alpha_{r}}{\alpha_{r}+\alpha_{b}},
\end{align}
%%%%%%%%%%%%%%%%%%%%%%%%%%%%%%%%%%%%%%%%%%%%%%
where $\eta_{abs}$ is the absorption efficiency coefficient. Equation ~(\ref{Eq:Netcoolingdens-3}) demonstrates several aspects of the laser cooling condition. In order for net cooling to be observed, the ASF cooling must dominate the heating terms, in turn yielding a negative net power density, $Q<0$. The cooling engine works in the long-wavelength tail of the absorption cross-section, where the pump and mean fluorescence wavelengths are very close each other, such that the ratio of $\lambda_{p} / \lambda_{f}$ is slightly greater than 1. So, in order for $Q$ to be negative, the product $\eta_{ext}\eta_{abs}$ must be very close to unity, implying the host material is of high purity ~\cite{epstein2010optical} . Cooling-grade materials may be defined by an external quantum efficiency close to 1, $\eta_{ext}>0.98$, and a background absorption much less than the resonant absorption, $\alpha_{b}/\alpha_{r}<0.01$~\cite{epstein2010optical}. The necessity of ultra pure hosts imposes a tight constraint on the diversity of suitable materials. 

%%%%%%%%%%%%%%%%%%%%%%%%%%%%%%%%%%%%%%%%%%%%%%%%%%%%%%%%%%%%%%%%%%%%%%%%%%%%%%%%%%%%%%%%%%%%%
\begin{figure}[h!]
\centering  
    \includegraphics[width=3 in]{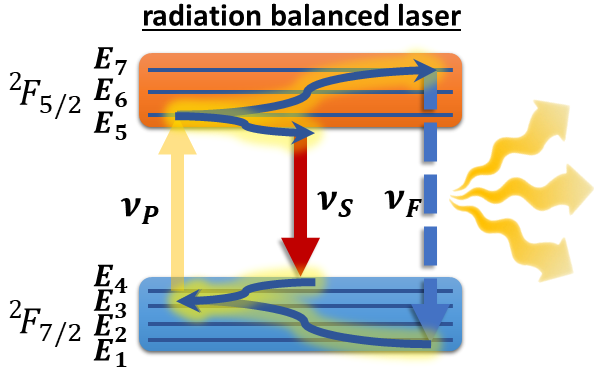}
\caption{Schematic diagram of radiation balancing engine. The heat generated by the quantum defect ($q_{qd}\propto (1-\frac{\nu_s}{\nu_p})$) is offset simultaneously by the heat extraction from anti-Stokes fluorescence cooling $(q_{f}\propto (1-\frac{\nu_f}{\nu_p}))$, $q=q_{qd}+q_{f}\approx 0.0$.}
\label{Fig:RBL-Schem}
\end{figure}
%%%%%%%%%%%%%%%%%%%%%%%%%%%%%%%%%%%%%%%%%%%%%%%%%%%%%%%%%%%%%%%%%%%%%%%%%%%%%%%%%%%%%%%%%%
\section{Radiation Balanced Laser Modeling}
 Figure \ref{Fig:RBL-Schem} concisely describes the idea of radiation balancing. Suppose the gain material of a laser is cooling-grade. Noting that signal amplification in any laser or amplifier is always associated with heat generation, let us imagine that the gain medium can simultaneously exhibit ASF cooling. It will be shown that with careful considerations, the heat generation originating from the quantum defect or other impurities in the laser can be offset by ASF cooling~\cite{bowman1999lasers,bowman2010minimizing,yang2019radiation}.
 The subtle balancing condition between different laser or amplifier parameters plays a critical role in radiation balancing, which is why S.~Bowman coined the term "Radiation Balanced Laser". Here, we will describe radiation balancing and the necessary conditions required for it.   

Unlike the case of a laser cooler, for a radiation balanced laser or amplifier, the signal amplification's contribution to the total heat must be taken into account. Considering the contribution of the signal amplification, the total heat power density can be described by
%%%%%%%%%%%%%%%%%%%%%%%%%%%%%%%%%%%%%%%%%%%%%%%%%%%%%%%%%%%%%%%%%%%%%%%%%%%%%%%%%%%
\begin{align}
\label{Eq:NetHeatDensRBL-1}
Q(\lambda_{p},\lambda_{s})&=[N_{0}-\frac{N_{2}}{\beta_{p}}]\sigma_{a}(\lambda_{p}) I_{p}+
[N_{0}-\frac{N_{2}}{\beta_{s}}]\sigma_{a}(\lambda_{s}) I_{s}-\frac{ h c}{\lambda_{f}\tau_{r}} N_{2}\\
\nonumber
&+\alpha_{b} I_{p}+\alpha_{b} I_{s},
\end{align}
%%%%%%%%%%%%%%%%%%%%%%%%%%%%%%%%%%%%%%%%%%%%%%%%%%%%%%%%%%%%%%%%%%%%%%%%%%%%%%%%%%%%%%%%%%%%%%%%%%
where the second term here represents the contribution of the signal amplification.

Decomposing $N_{2}$ into two parts,
\begin{align}
    \label{n2decomp}
    N_{2}=\frac{\beta_{p}i_{p}}{1+i_{p}+i_{s}}N_{0}+\frac{\beta_{s}i_{s}}{1+i_{p}+i_{s}}N_{0},
\end{align}
permits Equation~(\ref{Eq:NetHeatDensRBL-1}) to be re-written as 
%%%%%%%%%%%%%%%%%%%%%%%%%%%%%%%%%%%%%%%%%%%%%%%%%%%%%%%%%%%%%%%%%%%%%%%%%%%%%%%%%%%
\begin{align}
\label{Eq:NetHeatDensRBL-2}
Q(\lambda_{p},\lambda_{s})&=[N_{0}-\frac{N_{2}}{\beta_{p}}]\sigma_{a}(\lambda_{p}) I_{p}-\frac{hc}{\lambda_{f}\tau_{r}}[\frac{\beta_{p}i_{p}}{1+i_{p}+i_{s}}N_{0}]\nonumber\\
&+[N_{0}-\frac{N_{2}}{\beta_{s}}]\sigma_{a}(\lambda_{s}) I_{s}-\frac{hc}{\lambda_{f}\tau_{r}}[\frac{\beta_{s}i_{s}}{1+i_{p}+i_{s}}N_{0}] 
\nonumber\\
&+\alpha_{b} I_{p}+\alpha_{b} I_{s}. 
\end{align} 
%%%%%%%%%%%%%%%%%%%%%%%%%%%%%%%%%%%%%%%%%%%%%%%%%%%%%%%%%%%%%%%%%%%%%%%%%%%%%%%%%%%%%%%%%%%%%%%%%%

With some simplifications, Eq.~(\ref{Eq:NetHeatDensRBL-2}) takes a new form, 
%%%%%%%%%%%%%%%%%%%%%%%%%%%%%%%%%%%%%%%%%%%%%%%%%%%%%%%%%%%%%%%%%%%%%%%%%%%%%%%%%%%%%%%%%%%%%%%%%%%%
\begin{align}
\label{Eq:NetHeatDensRBL-3}
Q(\lambda_{p},\lambda_{s})&=\alpha_{r}(\lambda_{p},\lambda_{s})I_{p}\big(1-\eta_{ext}\frac{\lambda_{p}}{\lambda_{f}}\big)- \gamma_{s}(\lambda_{p},\lambda_{s})I_{s}\big(1-\eta_{ext}\frac{\lambda_{s}}{\lambda_{f}}\big)\nonumber\\
&+\alpha_{b}(I_{p}+I_{s}),
\end{align}
%%%%%%%%%%%%%%%%%%%%%%%%%%%%%%%%%%%%%%%%%%%%%%%%%%%%%%%%%%%%%%%%%%%%%%%%%%%%%%%%%%%%%%%%%%%%%%%%%%
where $\gamma_{s}$ (signal gain) and $\alpha_{r}$ (resonant absorption coefficient) are defined by
%%%%%%%%%%%%%%%%%%%%%%%%%%%%%%%%%%%%%%%%%%%%%%%%%%%%%%%%%%%%%%%%%%%%
\begin{align}
\label{Eq:SignalGain}
&\gamma_{s}(\lambda_{p},\lambda_{s})=\alpha_{0}(\lambda_{s})\Big[\frac{i_{p}\frac{\beta_{p}}{\beta_{ps}}-1}{1+i_{p}+i_{s}}\Big],\nonumber\\
&\alpha_{r}(\lambda_{p},\lambda_{s})= \alpha_{0}(\lambda_{p})\Big[\frac{i_{s}\frac{\beta_{s}}{\beta_{ps}}+1}{1+i_{p}+i_{s}}\Big].
\end{align}
%%%%%%%%%%%%%%%%%%%%%%%%%%%%%%%%%%%%%%%%%%%%%%%%%%%%%%%%%%%%%%%%%%%%%%%%%%%%%%%%%%%%%%%%%%%%%%%%%%

Neglecting the background absorption, Eq ~(\ref{Eq:NetHeatDensRBL-3}) finally can be re-written in the final form of
%%%%%%%%%%%%%%%%%%%%%%%%%%%%%%%%%%%%%%%%%%%%%%%%%%%%%%%%%%%%%%%%%%%%%%%%%%%%%%%%%%%%%%%%%%%%%%%%%%%%
\begin{align}
\label{Eq:NetHeatDensRBL-4}
Q(\lambda_{p},\lambda_{s})&=Q_{0}\frac{i_{p}i_{s}}{1+i_{p}+i_{s}}[1-\frac{i_{s}^{min}}{i_{s}}-\frac{i_{p}^{min}}{i_{p}}],\\
Q_{0}&=(\beta_{p}-\beta_{s})\frac{hc}{\tau_{r}}(\frac{1}{\lambda_{p}}-\frac{1}{\lambda_{s}})N_{0},\nonumber\\
i_{p}^{min}&=\frac{I_{p}^{min}}{I_{sat}(\lambda_{p})}=\frac{\lambda_{p}}{\lambda_{f}^{*}}\big(\frac{\lambda_{s}-\lambda_{f}^{*}}{\lambda_{s}-\lambda_{p}}\big)\frac{\beta_{ps}}{\beta_{p}},\\
i_{s}^{min}&=\frac{I_{s}^{min}}{I_{sat}(\lambda_{s})}=\frac{\lambda_{s}}{\lambda_{f}^{*}}\big(\frac{\lambda_{p}-\lambda_{f}^{*}}{\lambda_{s}-\lambda_{p}}\big) \frac{\beta_{ps}}{\beta_{s}}\nonumber,
\end{align}
%%%%%%%%%%%%%%%%%%%%%%%%%%%%%%%%%%%%%%%%%%%%%%%%%%%%%%%%%%%%%%%%%%%%%%%%%%%%%%%%%%%%%%%%%%%%%%%%%%
where $\lambda_{f}^{*}=\eta_{ext}\lambda_{f}$ is the crossing wavelength, beyond which no optical cooling is possible regardless of the material quality. 

Equation ~(\ref{Eq:NetHeatDensRBL-4}) shows that in order for the generated heat to vanish, ($Q=0$), the following condition, the radiation-balancing condition, has to be satisfied,
%%%%%%%%%%%%%%%%%%%%%%%%%%%%%%%%%%%%%%%%%%%%%%%%%%%%%%%%%%%%%%%%%%%%
\begin{align}
\label{Eq:RBLCond}
1-\frac{i_{s}^{min}}{i_{s}}-\frac{i_{p}^{min}}{i_{p}}=0.
\end{align}
%%%%%%%%%%%%%%%%%%%%%%%%%%%%%%%%%%%%%%%%%%%%%%%%%%%%%%%%%%%%%%%%%
%%%%%%%%%%%%%%%%%%%%%%%%%%%%%%%%%%%%%%%%%%%%%%%%%%%%%%%%%%%%%%%%%%%%%%%%%%%%%

%%%%%%%%%%%%%%%%%%%%%%%%%%%%%%%%
\begin{figure}[H]
\centering
\includegraphics[width=3.3 in]{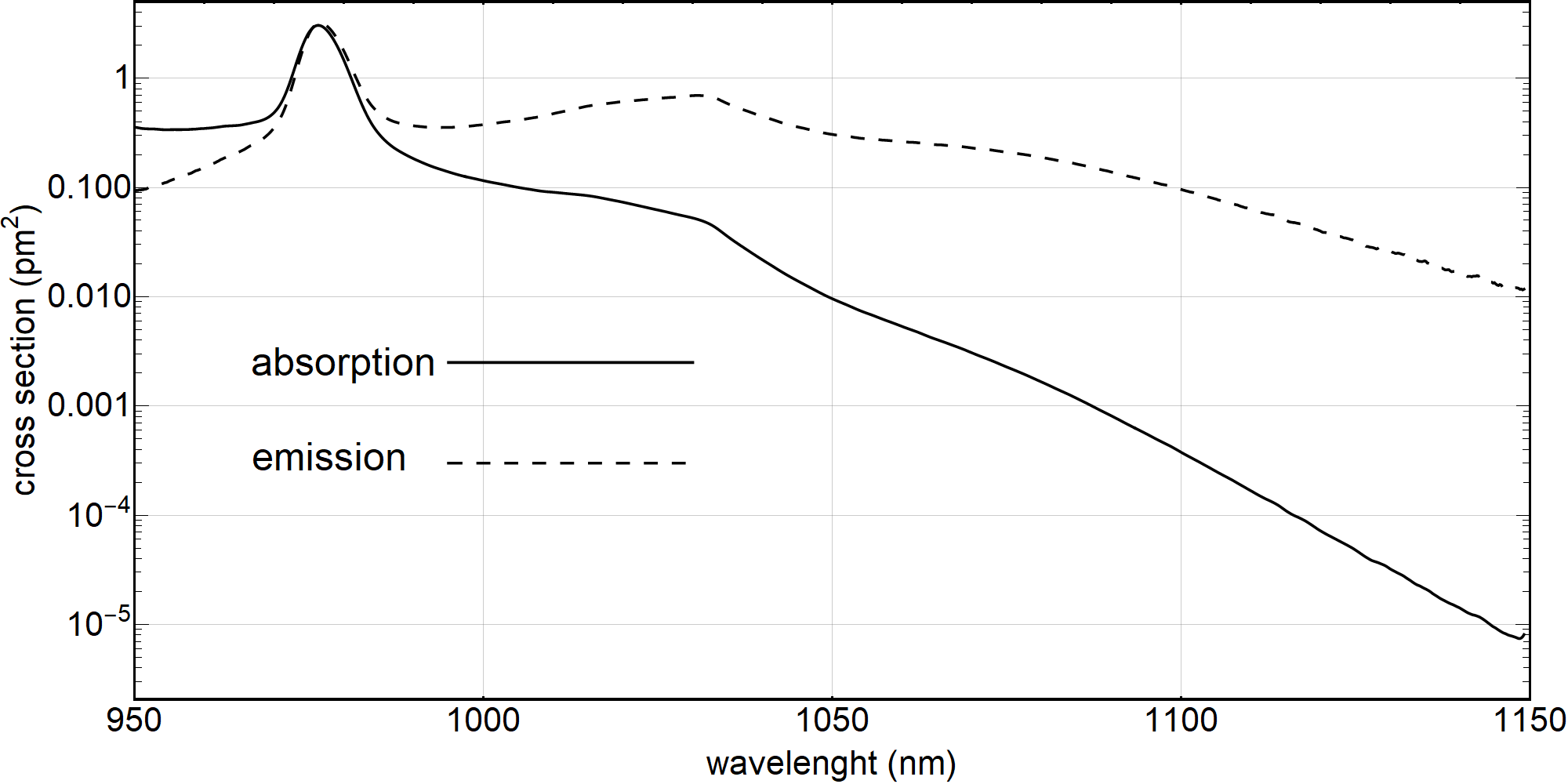}
\caption{Yb:Silica emission and absorption cross-sections }
\label{Fig:absem}
\end{figure}
%%%%%%%%%%%%%%%%%%%%%%%%%%%%%%%%%%%%%%

 It is clear that as the pump intensity becomes large ($i_{p}\to \infty$), the signal intensity asymptotically approaches a constant value of $i_{s}=i_{s}^{min}$. The same is also true for the pump intensity, if $i_{s}\to \infty$ then $i_{p}=i_{p}^{min}$. The relationship which is imposed by the radiation balancing condition on the pump and signal intensities is very different from the relationship that one can find in a typical laser or amplifier, for which the pump and signal have a linear relationship ($i_{p}\approx i_{s}$)~\cite{bowman1999lasers}.   
%%%%%%%%%%%%%%%%%%%%%%%%%%%%%%%%%%%%%%%%%%%%%%%%%%%%%%%%%%%%%%%%%%%%%%%%%%%%%
Due to the mutual dependence of pump and signal in fiber lasers, Eq.~(\ref{Eq:RBLCond}) can not be applied to fiber designs. For a cooling-grade material, we can reasonably assume the background absorption is negligible in comparison with the other terms in Eq.~(\ref{Eq:NetHeatDensRBL-1}). This equation can then be reduced to 
\begin{align}
\label{Eq:NetHeatDensRBL-1-lowabs}
Q(\lambda_{p},\lambda_{s})&=[N_{0}-\frac{N_{2}}{\beta_{p}}]\sigma_{a}(\lambda_{p}) I_{p}+
[N_{0}-\frac{N_{2}}{\beta_{s}}]\sigma_{a}(\lambda_{s}) I_{s}-\frac{ h c}{\lambda_{f}\tau_{r}} N_{2}\\
\nonumber.
\end{align}
In our previous work, we found an analytical solution for pump and signal in high-power fiber lasers. Here, our goal is to use our analytical solution to find an analytical equation for the heat power density.
We employ the notation of S. Bowman~\cite{bowman1999lasers} where 
\begin{align}
I_p(z)\,:=\,\frac{\Gamma_p ( P_p^+(z)+P_p^-(z))}{A},\\
I_s(z)\,:=\,\frac{\Gamma_s ( P_s^+(z)+P_s^-(z))}{A},
\end{align}
and 
\begin{align}
\beta_p\,:=\,\frac{\sigma^{a}_p}{\sigma^{a}_p + \sigma^{e}_p}=\frac{\sigma^{a}_p}{\sigma^{ae}_p} ,\\
\beta_s\,:=\,\frac{\sigma^{a}_s}{\sigma^{a}_s + \sigma^{e}_s}=\frac{\sigma^{a}_s}{\sigma^{ae}_s} .
\end{align}
For  simplicity,  we  assume  that the  pump  reflection  from  the  second  mirror is  negligible. By implementing our earlier notation Eq.~(\ref{Eq:NetHeatDensRBL-1-lowabs}) can be expressed as 
 \begin{align}
\label{Eq:NetHeatDensRBL-1-lowabs-new}
&Q(\lambda_{p},\lambda_{s})=[N_{0}-\frac{N_{2} \sigma^{ae}_p }{\sigma^{a}_p}]\sigma^{a}_p \frac{\Gamma_p ( P_p^+(z))}{A}\\\nonumber
&+[N_{0}-\frac{N_{2}\sigma^{ae}_s}{\sigma^{a}_s}]\sigma^{a}_s \frac{\Gamma_s ( P_s^+(z)+P_s^-(z))}{A}-\frac{ h c}{\lambda_{f}\tau_{r}} N_{2}.\\
\nonumber
\end{align}
%%%%%%%%%%%%%%%%%%%%%%%%%%%%%%%%%%%%%%
\begin{table}[H]
\centering
\caption{\bf YDCFL parameters}
\scalebox{0.8}{
\begin{tabular}{ccc}
 \hline
 Symbol & Parameter & Value\\
 \hline
$A$ & Core area & $5.0 \times 10^{-11}~m^2$ \\
$\Gamma_s$ & Signal power filling factor & 0.82\\
$\Gamma_p$ & Pump power filling factor & $1.2 \times 10^{-3}$\\
$N_0$ & $Yb^{3+}$ concentration & $6.0 \times 10^{25}~m^{-3}$\\
$\tau$ & Radiative lifetime & $1.0 ms$\\

$\lambda_p$ & Pump wavelength & $1030~nm$\\
$\lambda_s$ & Signal wavelength & $1090~nm$\\
$R_1$ & First reflector & $99 \%$\\
$R_2$ & Second reflector & $4 \%$\\
$L$ & Fiber length & $35~m$\\
$P_p^+(0)$ & Pump power & $40~W$\\
\end{tabular}}
  \label{tab:values}
\end{table} 
%%%%%%%%%%%%%%%%%%%%%%%%%%%%%%%%%%%%%
Based on our previous work outlining an analytical equation for signal and pump powers, $N_2(z)$ can be expressed as

\begin{align}
\label{n2}
\dfrac{N_2(z)}{N}=\dfrac{\dfrac{\Gamma_p \sigma^a_p P_p^+(z)}{h \nu_p A}+\dfrac{\Gamma_s \sigma^a_s (P_s^+(z)+P_s^-(z))}{h \nu_s A}}{\dfrac{\Gamma_p \sigma^{ae}_p P_p^+(z)}{h \nu_p A}+\dfrac{1}{\tau}+\dfrac{\Gamma_s \sigma^{ae}_s (P_s^+(z)+P_s^-(z))}{h \nu_s A}},
\end{align}
The solution for the pump propagation can be written as 
\begin{align}\label{ppumpprop}
P_p^+(z)=P_p^+(0) e^{- \alpha z},
\end{align}
where we have defined $\alpha$ according to
\begin{align}
\alpha\,:=\,(\Gamma_p \sigma^a_p N + \alpha_p).
\end{align}
The solution for the signal propagation is given by
\begin{align}
\begin{cases}\label{twoeq}
P_s^+(z) =-b + \sqrt{b^2 + R_1 {P_s^-(0)}^2},\\
P_s^-(z) =+b + \sqrt{b^2 + R_1 {P_s^-(0)}^2}.
\end{cases}
\end{align}
where
\begin{align}
\label{ps0app}
&P_s^-(0)=\frac{\sqrt{R_2}}{\left[\sqrt{R_1} (1-R_2) + \sqrt{R_2}(1-R_1) \right]}\\\nonumber
&\times \Bigg[ (\frac{\nu_s}{\nu_p})P_p^+(0) (1-\frac{\alpha_p}{\alpha}) (1-e^{-\alpha L}) \\\nonumber 
&- \frac{h \nu_s A}{\tau} (\frac{\ln \frac{1}{\sqrt{R_1 R_2}} +(\Gamma_s \sigma^a_s N + \alpha_s ) L}{\Gamma_s (\sigma^e_s + \sigma^a_s)}) \Bigg].
\end{align}
and  
\begin{align}
\label{quadone}
&b:= \frac{(1-R_1)P_s^-(0)}{2} - \frac{\nu_s}{2\nu_p} P_p^+(0) (1-\frac{\alpha_p}{\alpha}) (1-e^{-\alpha z})\\ 
\nonumber
&\quad+\frac{h \nu_s A \overline{N_2} z}{2 \tau}.\\\nonumber
\end{align}
where $\overline{N_2}$ depends only on the known laser parameters giving
\begin{align}\label{navg}
\overline{N_2} = \frac{\frac{1}{L} \ln \frac{1}{\sqrt{R_1 R_2}} +(\Gamma_s \sigma^a_s N + \alpha_s )}{\Gamma_s (\sigma^e_s + \sigma^a_s)}.
\end{align}

%%%%%%%%%%%%%%%%%%%%%%%%%%%%%%%%%%%%%%%%%%%

%%%%%%%%%%%%%%%%%%%%%%%%%%%%%%%%%
\begin{figure}[H]
\centering
\includegraphics[width=3.3 in]{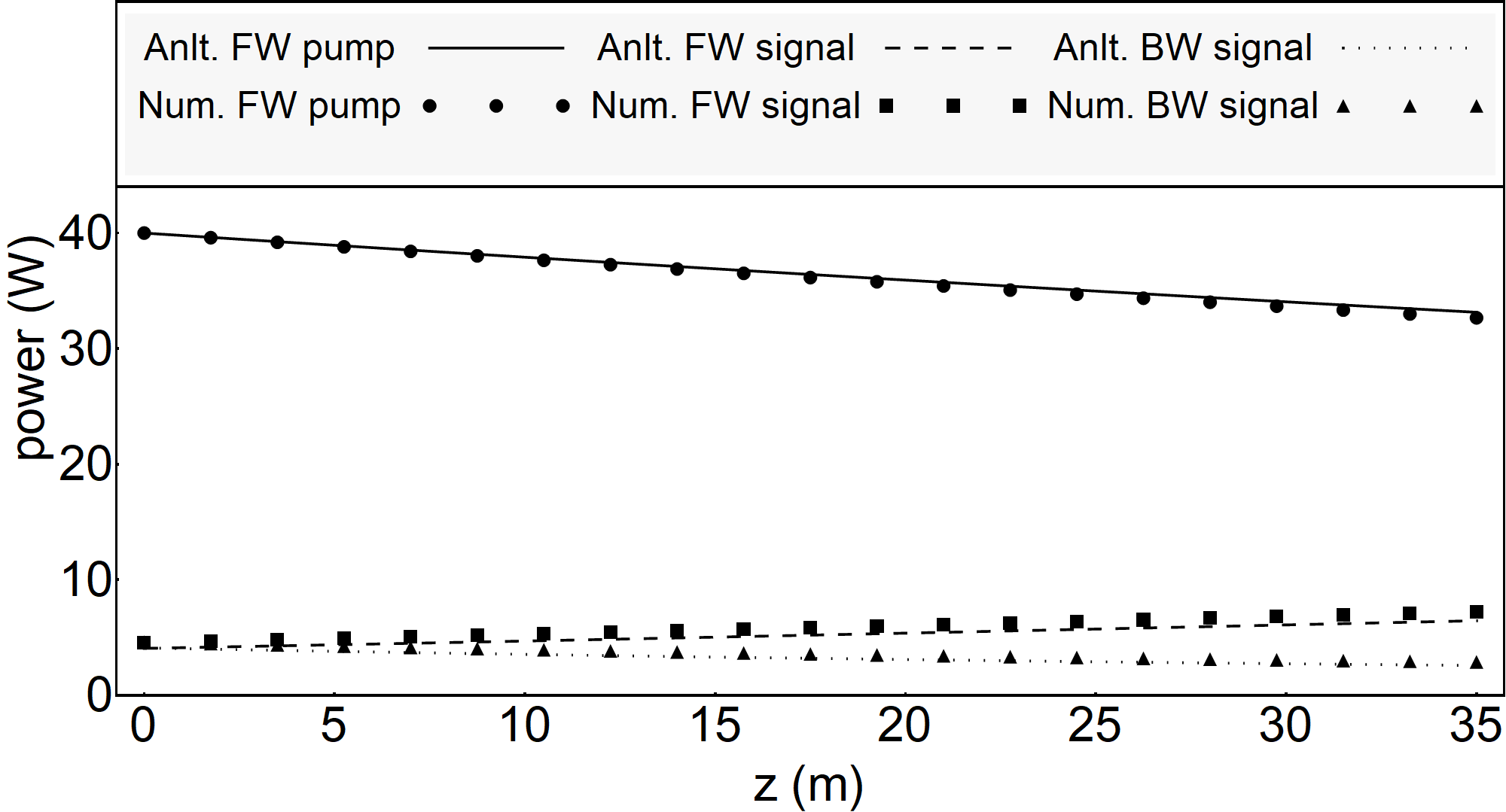}
\caption{Comparison of the propagation of the analytical forward pump (Anlt. FW pump), analytical forward signal (Anlt. FW signal), and analytical backward signal (Anlt. BW signal) with their corresponding numerical counterparts, the exact numerical forward pump (Num. FW pump), exact numerical forward signal (Num. FW signal), and exact numerical backward signal (Num. BW signal). Calculations were carried out using the set of parameter values represented in Table~\ref{tab:values}.}
\label{Fig:anaexac}
\end{figure}
%%%%%%%%%%%%%%%%%%%%%%%%%%%%%%%%%%%%%%%%

%%%%%%%%%%%%%%%%%%%%%%%%%%%%%%%%%
\begin{figure}[H]
\centering
\includegraphics[width=3.3 in]{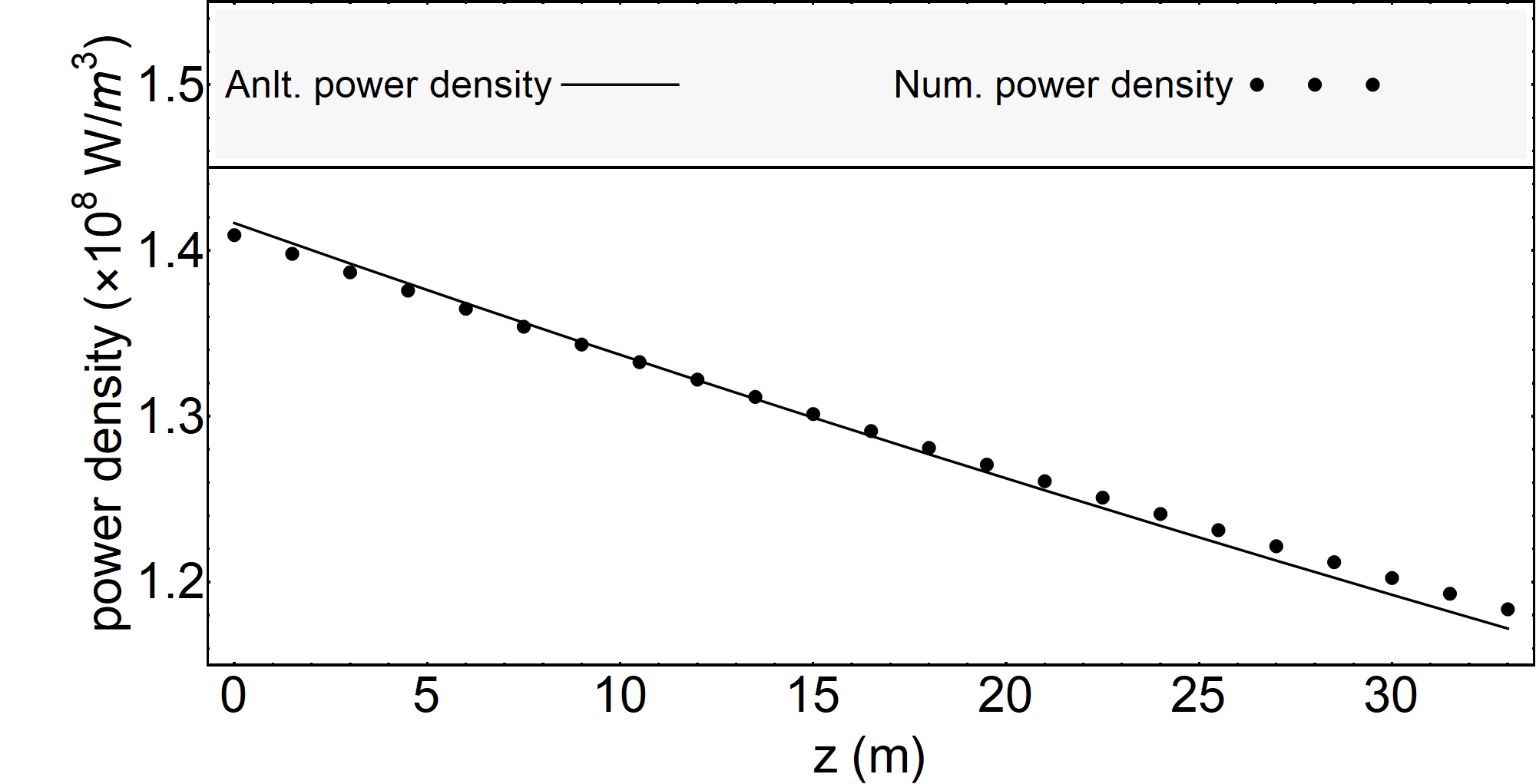}
\caption{Comparison of the analytical (anlt.) and numerical (num.) power densities. Calculations were carried out using the set of parameter values represented in  Table~\ref{tab:values}.}
\label{Fig:pdenscomp}
\end{figure}
%%%%%%%%%%%%%%%%%%%%%%%%%%%%%%%%%%%%%%%%

\section{Conclusion}
%%%%%%%%%%%%%%%%%%%%%%%%%%%%%%%%%%%%%%%%%%%%%%%%%%%%%%%%%%%%%%%%%%%%%%%%%%%%%%%%%%%%
%%%%%%%%%%%%%%%%%%%%%%%%%%%%%%%%%%%%%%%%%%%%%%%%%%%%%%%%%%%%%%%%%%%%%%%%%%%%%%%%%%%%
In summary, we presented an analytical solution for radiation balanced YDCFL. The results are in excellent agreement with the direct numerical solution. This study is important because calculating the pump and signal powers in a fiber laser in order to design a radiation balanced YDCFL using a numerical solution involves iterative solutions of coupled differential equations, which can be time-consuming. An analytical solution is especially
beneficial for multi-parameter optimization in designing a radiation balanced YDCFL. The presence of an analytical solution can significantly speed up optimization algorithms that target radiation balancing situations. The optimization parameters may include the choice of the fiber geometry (primarily the length), active ion concentration, and mirror reflectivities. In addition to the material parameters, since addressing RBL design for a fiber~\cite{Bowman,Mobini:18} requires consideration of pump and signal wavelengths, numerical optimization is a time-consuming and therefore economically expensive task~\cite{Peysokhan:20}. This study will pave the way to designing radiation-balanced YDCFLs over a significant design parameter space.

\section*{Acknowledgments}
%%%%%%%%%%%%%%%%%%%%%%%
%%%%%%%%%%%%%%%%%%%%%%%
This material is based upon work supported by the Air Force Office of Scientific Research under award number FA9550-16-1-0362 titled Multidisciplinary Approaches to Radiation Balanced Lasers (MARBLE).
\section*{Disclosures}
Disclosures. The authors declare no conflicts of interest.
% Bibliography
\bibliography{sample}

% Full bibliography added automatically for Optics Letters submissions; the following line will simply be ignored if submitting to other journals.
% Note that this extra page will not count against page length
\bibliographyfullrefs{sample}

%Manual citation list
%\begin{thebibliography}{1}
%\bibitem{Zhang:14}
%Y.~Zhang, S.~Qiao, L.~Sun, Q.~W. Shi, W.~Huang, %L.~Li, and Z.~Yang,
 % \enquote{Photoinduced active terahertz metamaterials with nanostructured
  %vanadium dioxide film deposited by sol-gel method,} Opt. Express \textbf{22},
  %11070--11078 (2014).
%\end{thebibliography}

% Please include bios and photos of all authors for aop articles
\ifthenelse{\equal{\journalref}{aop}}{%
\section*{Author Biographies}
\begingroup
\setlength\intextsep{0pt}
\begin{minipage}[t][6.3cm][t]{1.0\textwidth} % Adjust height [6.3cm] as required for separation of bio photos.
  \begin{wrapfigure}{L}{0.25\textwidth}
    \includegraphics[width=0.25\textwidth]{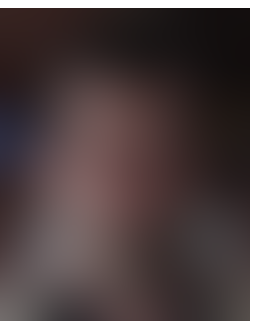}
  \end{wrapfigure}
  \noindent
  {\bfseries John Smith} received his BSc (Mathematics) in 2000 from The University of Maryland. His research interests include lasers and optics.
\end{minipage}
\begin{minipage}{1.0\textwidth}
  \begin{wrapfigure}{L}{0.25\textwidth}
    \includegraphics[width=0.25\textwidth]{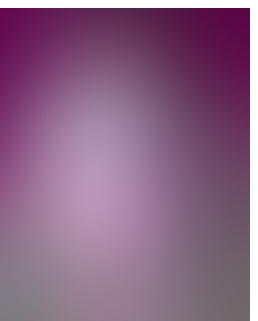}
  \end{wrapfigure}
  \noindent
  {\bfseries Alice Smith} also received her BSc (Mathematics) in 2000 from The University of Maryland. Her research interests also include lasers and optics.
\end{minipage}
\endgroup
}{}

\end{document}